\documentclass{article}

\usepackage{natbib}
\usepackage{graphicx}
\usepackage{float}
\usepackage{placeins}
\usepackage{adjustbox}
\usepackage{booktabs}
\usepackage[sglblindworkshop, final]{neurips_2025}
\usepackage[utf8]{inputenc} % allow utf-8 input
\usepackage[T1]{fontenc}    % use 8-bit T1 fonts
\usepackage{hyperref}       % hyperlinks
\usepackage{url}            % simple URL typesetting
\usepackage{booktabs}       % professional-quality tables
\usepackage{amsfonts}       % blackboard math symbols
\usepackage{nicefrac}       % compact symbols for 1/2, etc.
\usepackage{microtype}      % microtypography
\usepackage{xcolor}         % colors
\usepackage{enumitem}
\title{DAWZY: A New Addition to AI powered  "Human in the Loop" Music Co-creation}
\workshoptitle{AI for Music}

\author{
  \begin{tabular}{ccc}
    \textbf{Aaron C Elkins} & \textbf{Sanchit Singh} & \textbf{Adrian Kieback} \\
    \normalfont San Diego State University & \normalfont San Diego State University & \normalfont San Diego State University \\
    \texttt{aelkins@sdsu.edu} & \texttt{ssingh1949@sdsu.edu} & \texttt{akieback@sdsu.edu} \\[1.5em]
    \textbf{Sawyer Blankenship} & \textbf{Uyiosa Philip Amadasun} & \textbf{Aman Chadha}\thanks{Work done outside role at Amazon.}\\
    \normalfont San Diego State University & \normalfont San Diego State University & \normalfont San Diego State University \\
    \texttt{sablankenship@ucdavis.edu} & \texttt{uamadasun@sdsu.edu} & \texttt{hi@aman.ai}
  \end{tabular}
}

\begin{document}
\raggedbottom

\maketitle

\begin{abstract}
% Digital Audio Workstations (DAWs) give creators fine-grained control, but the gap between high-level intent (e.g., “warm the vocals”) and low-level operations (track selection, FX insertion, parameter scaling, routing) remains a barrier to flow. While recent LLM- and tool-based systems show promise, many either operate outside the DAW or rely on brittle GUI macros, offering limited state grounding, weak safety guarantees, and little support for learning. We present \textit{DAWZY}, an open-source assistant that translates natural-language requests (text/voice/hum) into reversible ReaScript actions in REAPER. \textit{DAWZY} contributes to the Music-in-AI ecosystem by (i) formalizing a minimal-GUI, voice-first interaction that preserves the DAW as the creative hub; (ii) introducing LLM-based code generation as a novel way to significantly reduce the time users spend familiarizing themselves with large interfaces, (iii) providing explain-as-you-go rationales that turn edits into micro-tutorials. It addresses core challenges—linguistic ambiguity (intent confirmation), context grounding (fresh state queries before mutation), and safety and reversibility (atomic scripts with single-shot undo)—and demonstrates reliability on common production tasks. Code and a short demo are available online.
Digital Audio Workstations (DAWs) offer fine control, but mapping high-level intent (e.g., “warm the vocals”) to low-level edits breaks creative flow. Existing artificial intelligence (AI) music generators are typically one-shot, limiting opportunities for iterative development and human contribution. We present \textit{DAWZY}, an open-source assistant that turns natural-language (text/voice/hum) requests into reversible actions in REAPER. \textit{DAWZY} keeps the DAW as the creative hub with a minimal GUI and voice-first interface. Its LLM-based code generation replaces complex menus with a simple chat box, reducing time spent learning interfaces. {DAWZY} also uses three Model Context Protocol tools for live state queries, parameter adjustment, and AI beat generation. It maintains grounding by refreshing state before mutation; and ensures safety and reversibility with atomic scripts and undo. In evaluations, DAWZY performed reliably on common production tasks and was rated positively by users across Usability, Control, Learning, Collaboration, and Enjoyment. We show reliability on common production tasks; code and a short demo are available.
\footnote{\textbf{Resources:} \href{https://anonymous.4open.science/r/DAWZY-92BE/README.md}{Code (anonymous)} \quad
\href{https://www.youtube.com/watch?v=RQmCuYLkEDk}{Demo (anonymous)}}
\end{abstract}
\vspace{-17pt}
\section{Introduction}\label{sec:intro}
\vspace{-5pt}
Modern music production centers on Digital Audio Workstations (DAWs) \cite{10.5555/975278}, which democratize pro-quality creation but burden users with option overload that disrupts flow \citep{Kjus2024Platformization}. A gap persists between high-level intent (e.g., “make the vocals warmer”) and the low-level steps to realize it.

Prior work points to a path forward: mature DAW scripting (Ableton’s Max for Live/Live API \citep{AbletonMaxForLive}, REAPER’s ReaScript/JSFX \citep{ReaperReaScript}), co-creative agents inside production loops (e.g., Juice \citep{Juice2024}), fully generative systems largely outside fine-grained editing (e.g., Suno \citep{Suno}), advances in code generation \citep{Chen2021Codex}, and standardized tool invocation via the Model Context Protocol (MCP) \citep{AnthropicMCP,hou2025modelcontextprotocolmcp}.

We introduce \textit{DAWZY}, an open-source assistant that maps natural-language requests to precise, context-aware, reversible ReaScript actions in REAPER. DAWZY queries live session state, emits auditable edits, and favors a minimal-GUI, voice-first workflow with plain-language explanations to support learning. It primarily interacts with REAPER through LLM code generation. Ableton-MCP \citep{Ableton_MCP} is an open-source, similar but less powerful tool that allows LLMs to c the Ableton Live API. Related efforts (e.g., Mozart AI \citep{MozartAI}) explore closed-source adjacent ideas; DAWZY emphasizes open-source availability and ReaScript-specific reliability, complementing rather than replacing existing tools.

\paragraph{Primary Contributions}
\vspace{-5pt}
\begin{itemize}[leftmargin=*, itemsep=0.3em]
  \item \textbf{System design \& open-source prototype.} REAPER-targeted pipeline mapping natural language to safe, reversible ReaScript grounded in live state (Sec.~\ref{sec:arch}).
  \item \textbf{MCP tool suite.} Permissioned tools for state query, unit-consistent FX parameter adjustment, and AI beat generation; supports future cross-DAW portability (Sec.~\ref{subsec:processing}).
  \item \textbf{Minimal-GUI, voice-first interaction.} Natural-language control with buttons for common tasks(“start,” “stop,” “record,” “undo”) to reduce GUI micromanagement (Sec.~\ref{subsec:user}).
  \item \textbf{Explain-as-you-go pedagogy.} Plain-language rationales accompany each edit to support learning and auditability.
  \item \textbf{ReaScript-focused model adaptation.} Plan to fine-tune an open-source LLM for reliable REAPER code generation (Sec.~\ref{sec:conclusion}).
\end{itemize}

\section{DAWZY Architecture}\label{sec:arch}

DAWZY comprises three layers (Figure~\ref{fig:architecture}): \emph{User Interaction}, \emph{Processing}, and \emph{Execution}, which capture natural-language intent, interpret it, and translate it into precise DAW operations.

\subsection{User Interaction Layer}\label{subsec:user}
The User Interaction Layer is a minimal-GUI entry point for expressing intent via \textbf{text}, \textbf{speech}, or \textbf{humming}, mediated by an \textbf{Electron.js} app \citep{Electron}. Given the complexity of traditional DAW interfaces, DAWZY prioritizes direct, natural-language control to reduce GUI micromanagement.
(1) \textbf{Text} — Users type commands/questions in Electron; queries are forwarded as text. 
(2) \textbf{Speech} - Spoken commands are transcribed by \textbf{Whisper} \citep{Radford2022Whisper} and follow the same downstream path (hands-free).
(3) \textbf{Humming} - A “record hum” button captures sketches; a local \textbf{BasicPitch} pipeline \citep{Spotify_BasicPitch_About,2022_BittnerBRME_LightweightNoteTranscription_ICASSP} converts audio to MIDI, which is auto-imported into REAPER as a new track.

\subsection{Processing Layer}\label{subsec:processing}
The Processing Layer turns user input into context-aware DAW operations. Off-the-shelf LLM approaches often hallucinate commands, mis-index tracks/parameters, or ignore live state. DAWZY constrains behavior via a reliable LLM, and context grounding.

\begin{itemize}[leftmargin=*, itemsep=0.2em]
  \item \textbf{Electron gateway.} Routes all queries to the LLM and returns responses/confirmations; hummed audio is sent to the hum-to-MIDI pipeline.
  \item \textbf{LLM.} We use \textbf{OpenAI GPT-5} \citep{OpenAI_Introducing_GPT5_2025,OpenAI_GPT5_SystemCard_2025} to interpret intent, call MCP tools, and emit Lua ReaScript. Open-source baselines (e.g., Qwen3-Coder-480B-A35B-Instruct \citep{Qwen3_HF_Collection,qwen3technicalreport}) underperformed, frequently producing invalid indices when the full context (track/parameter mappings) was not considered; GPT-5 generated reliable edits.
  \item \textbf{Model Context Protocol (MCP).} Exposes DAW capabilities as explicit, permissioned functions between the LLM and REAPER:

    \begin{itemize}[leftmargin=1.2em, itemsep=0.15em]
      \item \textbf{State query.} Enumerates tracks, items, FX, and routing to ground edits in live session state and keep tool calls synchronized.
      \item \textbf{FX parameterization (\texttt{fxparam}).} Converts human units (dB, ms) to ReaScript slider ranges (e.g., 0–1, 0–4) to prevent scaling errors. Code generation failed here because the LLM could not reliably convert between units.
      \item \textbf{Beat generation.} Meta's \textbf{MusicGen-small (300M)}  model is run locally to create an audio waveform based on a text description \citep{Meta_MusicGen_Small_HF,copet2023simple}.
  \end{itemize}
  \item \textbf{Hum to MIDI.} The open-source \textbf{Spotify BasicPitch} model is run locally to convert hums into MIDI data \citep{Spotify_BasicPitch_About,Bittner2022BasicPitch}.
\end{itemize}

\begin{figure}[!t]
  \centering
  % Adjust trim values if you want to shave off white margins (L B R T, units can be mm)
    \includegraphics[width=0.85\linewidth, clip, trim=0 5 0 5]{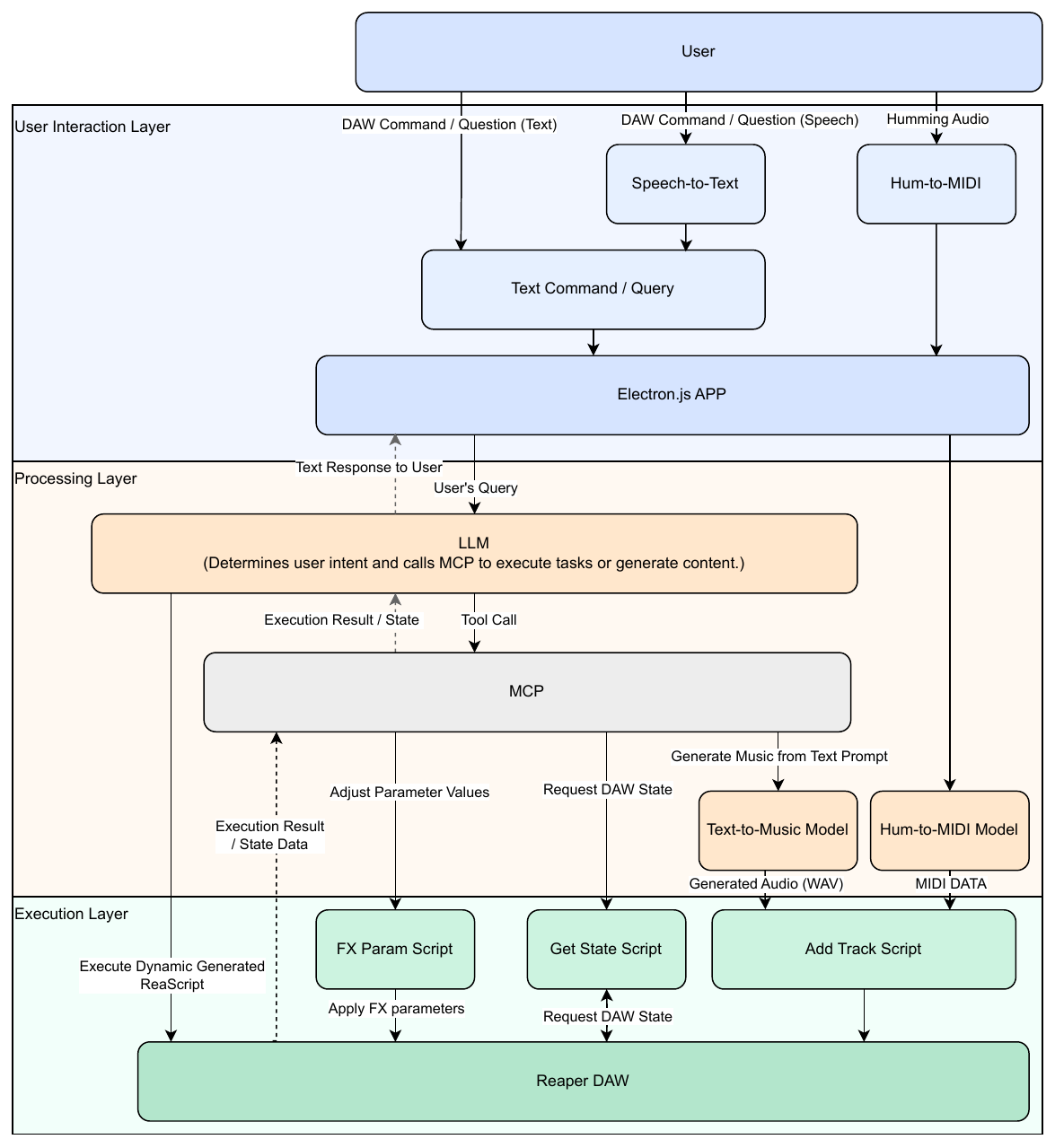}
    \vspace{2mm}
    \caption{\textbf{DAWZY Architecture.} User intent (text/speech/hum) flows through the Electron gateway to the LLM and MCP tools, then executes as reversible ReaScripts in REAPER. Rounded rectangles denote AI/MCP components; sharp rectangles denote DAW/runtime components; dashed arrows indicate data queries; solid arrows indicate state-changing actions.}
  \label{fig:architecture}

\end{figure}
% \subsection{Execution Layer}\label{subsec:execution}
% The Execution Layer (Figure~\ref{fig:architecture}) enacts operations inside REAPER in a safe and transparent manner. The main challenge is that even once user intent has been correctly interpreted, carrying it out inside a DAW can be fragile: actions must be applied in a way that is safe, reversible, and understandable to the user. 
% DAWZY addresses this by grounding all operations in REAPER’s live project state and 
% executing them through ReaScripts, ensuring that edits remain robust, 
% transparent, and easy to follow (cf. \textbf{System design \& prototype} and \textbf{Explain-as-you-go pedagogy} contributions).

% \begin{itemize}
%     \item \textbf{ReaScript actuation.} Dynamically generated Lua scripts from 
%     GPT 5 are executed via ReaPy, directly modifying the REAPER project state.
    
%     \item \textbf{Scoped tool scripts.} Dawzy provides specialized ReaScripts for:
%     \begin{itemize}
%         \item applying FX parameter changes,
%         \item summarizing REAPER’s project state,
%         \item and adding audio or MIDI files as new tracks.
%     \end{itemize}
    
%     \item \textbf{DAW integration.} REAPER remains the central creative hub: Dawzy 
%     extends rather than replaces the DAW workflow, grounding all automation in 
%     live project state rather than brittle GUI macros.
% \end{itemize}
\subsection{Execution Layer}\label{subsec:execution}
The Execution Layer (Figure~\ref{fig:architecture}) performs edits in REAPER safely, reversibly, and transparently by grounding actions in live project state. 
(1) \textbf{ReaScript actuation} - GPT-5 generates Lua that ReaPy executes to modify the project; changes are reversible.
(2) \textbf{Utility scripts} - Specialized scripts handle (i) FX parameter updates, (ii) project-state summaries, and (iii) audio/MIDI import as new tracks.
% \begin{itemize}[leftmargin=*, itemsep=0.2em]
%   \item \textbf{ReaScript actuation.} GPT-5 generates Lua that ReaPy executes to modify the project; changes are reversible.
%   \item \textbf{Utility scripts.} Specialized scripts handle (i) FX parameter updates, (ii) project-state summaries, and (iii) audio/MIDI import as new tracks.
% \end{itemize}

\setlength{\abovecaptionskip}{4pt}
\setlength{\belowcaptionskip}{2pt}
\setlength{\textfloatsep}{8pt}

\vspace{-10pt}
\section{Evaluation}\label{sec:eval}
\vspace{-5pt}
We evaluate DAWZY using both objective and qualitative performance tasks, as well as subjective user ratings.

\vspace{-5pt}

\subsection{Objective Evaluation}
\vspace{-5pt}

% To test reliability, we designed four reproducible tasks:
% (1) \textbf{Multi-instruction FX processing} — evaluating parameter control under compound instructions, 
% (2) \textbf{GUI navigation} — testing accessibility of DAW navigation, e.g., opening the FX browser through natural-language commands, 
% (3) \textbf{Workflow automation} — testing multi-step creative edits such as duplication, pitch-shifting, and blending, and 
% (4) \textbf{Educational interaction} — evaluating whether the system enriches interactions through pedagogy, e.g., explaining production concepts like compressor parameters in plain language. 
% Together, these tasks probe DAWZY’s reliability across parameter control, accessible navigation, multi-step automation, and natural-language pedagogy.
To test reliability, we designed four reproducible tasks:
(1) \textbf{Multi-instruction FX processing} — "Double the first track's volume, increase the decay, and set the attack to 10\,ms,"
(2) \textbf{GUI navigation} — "Open the FX browser for the first track," 
(3) \textbf{Workflow automation} — "Duplicate the first track, pitch it up one octave, and blend it in at 20\%," and
(4) \textbf{Educational interaction} — "What does attack time do in the second track's compressor?". Building on the four tasks, we ran 3 trials per task for 4 different LLMs. Open-source baselines (QWEN-480B, GPT-OSS-120B, GPT-OSS-20B) achieved only 25–50\% success, often failing due to hallucinated or invalid ReaScript functions and mis-indexed parameters. All models did however pass all 3 trials on the Education task.
% \begin{enumerate}[leftmargin=*, itemsep=2pt]
%   \item \textbf{Multi-Instruction FX Processing:} ``Double the first track's volume, increase the decay, and set the attack to 10\,ms.'' This task tests the LLMs ability to call our FX Parameter Adjustment MCP tool with the correct arguments.
%   \item \textbf{GUI Navigation:} ``Open the FX browser for the first track.'' This demonstrates DAWZY's capacity to interact with configuration windows through the ReaScript API.
%   GUI manipulation API functions are rarely used and sparsely documented. This tests whether the LLM has been trained on the entire API or just on popular functions.
%   \item \textbf{Creative workflow automation:} ``Duplicate the first track, pitch it up one octave, and blend it in at 20\%.'' This task tests the LLMs ability to generate effective code interacting with distinct REAPER components. Pitching up is achieved either by adjusting the detune parameter, adding a ReaPitch effect, or shifting each MIDI note. 
%   \item \textbf{Educational interaction:} ``What does attack time do in the second track's compressor?'' DAWZY provides explanation of compressor parameters using metaphor. It uses natural language to explain when to increase and decrease the parameter. No code should be written for this task. No MCP calls should be made.
% \end{enumerate}

\subsection{Ableton-MCP Comparison}
\vspace{-5pt}

We compared DAWZY and Ableton-MCP on three qualitative tasks: "Make notes slide into each other" (Wavy), "Make the track bouncy" (Bouncy), and "Make the track fade" (Fade). Success was determined by perceived audio changes or external validation (Google/ChatGPT). Both systems used Claude Sonnet 4.5. DAWZY (REAPER with custom prompt) encountered 3 execution errors on Fade but otherwise performed successfully. Ableton-MCP (Ableton Live via Claude Desktop) failed Wavy and Fade due to reported API limitations for modifying notes and volume. For Bouncy, it either incorrectly claimed success or created new segments instead of modifying existing ones. DAWZY's superior performance likely reflects Claude's pre-training on REAPER's publicly available documentation, whereas Ableton Live's API is newer and underrepresented in training data. Based of this result we expect similar results for open-source models like GPT-OSS-120B.

\begin{table}[h] 
\centering
\begin{tabular}{lcc} 
\toprule
Task & DAWZY & Ableton-MCP \\ 
\midrule 
Wavy & 3 & 0 \\ 
Bouncy & 1 & 0 \\ 
Fade & 0 & 0 \\ 
\midrule 
Success Rate & 44\% & 0\% \\ 
\bottomrule 
\end{tabular}
\vspace{2mm} 
\caption{\textbf{Task success comparison.} Scores denote successful trials out of three attempts per task.} 
\label{tab:Ableton-MCP comparison} 
\end{table}
\vspace{-13pt}

\subsection{Subjective Evaluation (MOS Test)}
We conducted a \textbf{Mean Opinion Score (MOS)} test with \textit{21} participants, who rated DAWZY's \textit{Enjoyment} as \(4.48\), \textit{Learning} as \(4.38\), \textit{Collaboration} as 4.29, \textit{Usability} as 4.14,
and \textit{Control} as 3.81 out of 5.
\vspace{-5pt}
\begin{figure}[H]
  \centering
  \includegraphics[width=0.8\textwidth]{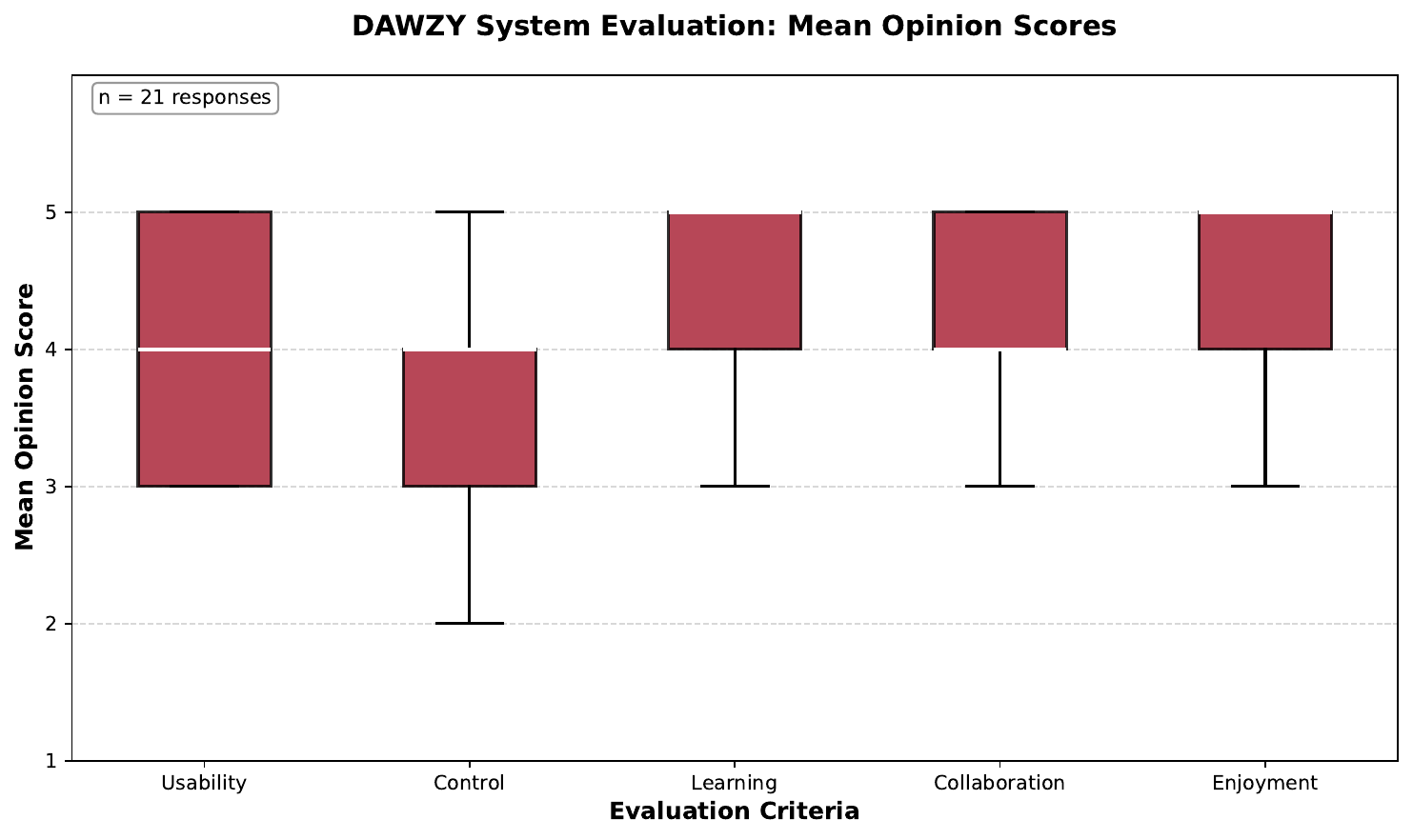}
  \caption{Mean Opinion Score (MOS) results for DAWZY (N=21). All categories scored above neutral (3).}
  \label{fig:mos-plot}
\end{figure}
\vspace{-15pt}
\section{Conclusion}\label{sec:conclusion}
\vspace{-7pt}
\textit{DAWZY} shows that natural-language control can enhance—not replace—human creativity in music software. By combining state tracking with context-aware code generation, it makes precise, reversible edits while keeping users in control. Current limitations stem from software-specific scripting languages and the need to adapt our system for each music program; our prototype focuses on core functions rather than advanced plugin features. As music software expands scripting capabilities \citep{AbletonLive12} and AI code generation improves \citep[p.~2]{Alenezi2025AISE}, we expect wider adoption in professional settings. Our key contribution is transparency: users can see and modify what the system does.  Next steps: (i) qualitative user studies, (ii) training or finetuning models specifically for more accurate music software scripting, and (iii) supporting scriptable DAWs beyond REAPER. We invite the community to try our demo and collaborate on this open framework for natural-language creative software control.
\begingroup
\small % 9pt is acceptable
\bibliographystyle{unsrtnat} % numeric, ordered by appearance
\bibliography{references}
\endgroup

\end{document}